\documentclass[12pt]{article}
\usepackage{indentfirst,graphicx}
\usepackage{cite}
\usepackage{doublespace}
\newcommand{\restrict}{{\cal M}}
\newcommand{\lift}{{\cal \mu}}
\begin{document}

\title{From molecular dynamics to coarse self-similar solutions: \\
A simple example using equation-free computation.}

\author{L. Chen${}^{1}$,
P. G. Debenedetti${}^{1}$,
C. W. Gear$^{1,2}$
and I.G. Kevrekidis${}^{1,3}$
\footnote{
Corresponding author, {\tt yannis@princeton.edu, tel (609) 258-2818,
FAX (609) 258-0211 }} \\
\small ${}^1$ Department of Chemical Engineering,
Princeton University, Princeton, NJ 08544; \\
\small ${}^2$ NEC Institute, retired;\\
\small ${}^3$ Program in Applied and Computational Mathematics,
Princeton University, Princeton, NJ 08544. \\
}

\date{\today}
\maketitle

\abstract{
In the context of the recently developed ``equation-free" approach to
the computer-assisted analysis of complex systems, we illustrate
the computation of coarsely self-similar solutions.
Dynamic renormalization and fixed point algorithms for the macroscopic
density dynamics are applied to the results of short bursts of
appropriately initialized molecular dynamics in a simple
diffusion simulation.
The approach holds promise for locating coarse self-similar
solutions and the corresponding exponents in a variety of multiscale
computational contexts.}


\newpage

{\noindent \bf \large 1. Introduction}

In contemporary scientific and engineering modeling we are often faced
with situations where the best system models are available at a {\it fine}
scale (e.g. atomistic, individual-based); yet we want to predict the
behavior of the system at a much more coarse-grained, macroscopic level.
In traditional modeling successful {\it closures} often allow us to
write models directly at the coarse-grained, macroscopic level at which
we want to model the behavior; typical examples include chemical kinetics
closures in terms of reactant concentrations for reactor modeling,
or Newton's law of viscosity in the
Navier-Stokes equations.
Often, however, such accurate closures are not available, and the
immense range of active scales (in space as well as time) precludes
the practical prediction of macroscopic behavior through direct
atomistic simulation.

The recently developed ``equation free" approach to coarse-grained
computer-assisted analysis of complex systems attempts to bridge this
enormous scale gap {\it when macroscopic equations conceptually exist,
but are not available in closed form}
\cite{TQK00,GKT02,MMK02,SGK03,LKG03,GLK03,GHK03,HK03}.
The approach constitutes a bridge between traditional continuum
numerical analysis and microscopic simulation.
The main idea is to consider the microscopic simulation as a {\it
computational experiment} which can be initialized and run at will.
If a coarse-grained, macroscopic equation were available, traditional
numerical tasks would involve repeated {\it evaluations} of the
equation, and of its functional or parametric derivatives, at various values
of the macroscopic state.
The idea is to substitute these {\it function evaluations} with short
bursts of appropriately initialized microscopic computational
experimentation, from the results of which the requisite numerical
quantities are {\it estimated}.
The quasi-continuum method of Phillips, Ortiz and coworkers
\cite{Ortiz}, as well as the optimal predictor approach of Chorin
and coworkers \cite{Chorin} embody many of these features; see
\cite{GHK03} for a discussion.
The approach can be successfully combined with matrix-free iterative
linear algebra to allow us to solve linear and nonlinear macroscopic
equations, perform ``coarse projective integration" as well as
additional tasks like controller design and optimization {\it without
ever writing down the macroscopic equations in closed form}.
This is a system identification based, ``closure on demand" approach.

Over the last few years we have demonstrated how to use this approach
for the accelerated simulation and bifurcation analysis of the
coarse-grained, expected behavior of kinetic Monte Carlo, Brownian
Dynamics, Molecular Dynamics and Lattice-Boltzmann microscopic
simulation codes,
\cite{TQK00,GKT02,MMK02,SGK03,LKG03,GLK03,GHK03,HK03,LB} as well as
the solution of {\it effective medium equations} \cite{Olof,Discrete}
for media with spatially varying properties.
The approach gives rise to two-level (conceptually possibly
multi-level) codes.
The ``inner code" is the best microscopic simulation of the phenomenon
at our disposal.
The outer code, the ``wrapper", is typically templated on traditional
continuum numerical analysis, and constitutes a protocol for {\it the
design} of microscopic simulations and for {\it processing their
results} towards a macroscopic modeling goal.
Accelerated simulation, the location of coarse fixed points and their
continuation and bifurcation analysis are typical such goals.

In many cases of interest, the macroscopic dynamics we explore do not
involve {\it stationary} solutions, but rather {\it traveling} or {\it
self-similar} solutions.
In this paper we will show how to use the basic coarse timestepper
methodology to construct dynamic renormalization algorithms for the
location of coarse self-similar solutions and the corresponding
similarity exponents by acting on a microscopic code directly.
Our illustrative example -- molecular diffusion in a thin two
dimensional domain -- is extremely simple, yet it does illustrate the
basic ingredients of the approach.
%%%%%
%%Y+L comment
% There is a real issue here about isothermal vs. constant
% energy - it should be OK but we should talk about it
%%%%%
%

The paper is organized as follows:
In the next Section we will briefly describe template-based dynamic
renormalization when macroscopic equations are explicitly available.
In Section 3 we summarize the main components of the coarse
timestepper, and, in particular, its modification for computation of
self-similar solutions as fixed points of the appropriate discrete
time map.
Section 4 presents our illustrative examples for both coarse dynamic and
coarse fixed point computations in a dynamic renormalization context.
We conclude with a brief summary and outline of the scope of the method and
some of the challenges we expect to arise in its wider application.

{\noindent \bf \large 2. Dynamic renormalization in a continuum context}

In problems with translational invariance, one often encounters
{\it traveling wave} solutions -- constant shape waves that move
in space at constant speed.
It is convenient (mathematically, computationally and practically)
to study these solutions, and the transient approach to them, in
a co-traveling frame.
In this frame the traveling solution appears stationary, and it is much
easier to study the transient approach to it and its stability
unencumbered by its constant motion.
Good techniques exist for computationally locating the translationally
invariant solution along with its traveling speed as a nonlinear
eigenvalue problem.
During transient simulation, however, the solutions both travel and approach
their ultimate, translationally invariant shape; the right speed for
``moving along" with such a transient solution may change from moment to
moment, and the best way to choose it is not transparent.
In a recent paper Rowley and Marsden \cite{Clancy} described a {\it
template based} approach that allows one to systematically recover
such an appropriate instantaneous speed; as the transient solution
asymptotically approaches the ultimate, translationally invariant,
traveling wave, the speed from the template-based algorithm
asymptotically approaches the correct traveling speed.

In a sequence of papers \cite{Betelu,Siettos,ClancyYannis} we have
shown how to adapt this approach from the computation of
translationally invariant solutions to the case of {\it scale
invariant} solutions -- that is, solutions of dynamic equations that evolve
across scales.
Self-similar solutions \cite{Barenblatt,Goldenfeld}
are an important such class of scale-invariant
solutions; in the same sense that it is practical to observe a
traveling solution in a co-traveling frame, self-similar solutions
are convenient to observe in a co-exploding or co-collapsing frame.
Consider the rather general form of partial differential equation
\begin{equation}\label{eq1}
u_t = D_x(u)
\end{equation}
where the operator $D$ satisfies the scaling relation
\begin{equation}\label{eq2}
D_x(Bf(x/A)) = A^aB^bD_y(f(y)), {\rm ~~~where~} y = x/A
\end{equation}
for $A>0, B>0$, and $a$ and $b$ are constants.
%
%{\bf ***}
We assume that there exists a
family of self-similar solutions,
\begin{equation}
u(x,t) = s^{\beta} U\left(\frac{x}{s^{\alpha}} \right);
\label{alpbet}
\end{equation}
where $s = |t-t^*|$ (for the case of
finite time blow-up at $t=t^*$) or $s = t$ otherwise, and $U$ satisfies the ODE,
\begin{equation}
\sigma \beta U-\alpha y U_y = D_y(U),
\label{Eq:ode}
\end{equation}
where $y = x/(s^\alpha)$ with $\sigma = sgn(t-t^*)$ and
\begin{equation}
\beta - 1 = a\alpha+b\beta.
\label{Eq:exp1}
\end{equation}
%%%
% Need to do something here ?
% !!!!!!!!!!!!!!!!!!!!!!!!!!!!!!!!!!!!!!!!!!!!
%(Note that
%there may be a one parameter family of solutions
%corresponding to different amplitudes, and that a different time
%snapshot of a self similar solution will be different, so gives a
%second degree of indeterminacy.)

%We assume that there exists a
%one-parameter family of self-similar solutions,
%\begin{equation}
%u(x,t) = s^{\beta} U\left(\frac{x}{Cs^{\alpha}} \right);
%\end{equation}
%where $C>0$ is
%{\bf *** isn't C a constant?  Also, it seems to have been taken as 1
%in the discussion below.  Why do we have it, it adds nothing but
%confusion (for this reader)?}
%the parameter, $s = |t-t^*|$ (for the case of
%finite time blowup at $t=t^*$), and $U$ satisfies the ODE,
%\begin{equation}
%\sigma C^{-a}(\beta U-\alpha y U_y) = D_y(U), {\rm ~~~where~} y = x/(Cs^\alpha)
%\label{Eq:ode}
%\end{equation}
%where $\sigma = sgn(t-t^*)$ and
%\begin{equation}
%\beta - 1 = a\alpha+b\beta.
%\label{Eq:exp1}
%\end{equation}

We are interested in locating the self-similar shape of the solution
as well as both similarity exponents $\alpha$ and $\beta$: one
more equation is needed for this latter task.
Starting with the general scaling
\begin{equation}
u(x,t) = B(s) w(\frac{x}{A(s)}, \tau(s))
\end{equation}
where $A$, $B$ and $\tau$ are unknown functions, and setting $\tau_s(s)
= \sigma A^{a}B^{b-1}$, the PDE becomes
\begin{equation}
w_{\tau}+\frac{B_{\tau}}{B}w - \frac{A_{\tau}}{A}yw_y =
D_y(w), {\rm ~~~where~} y = x/A(\tau).
\label{Eq:AB}
\end{equation}
%\begin{eqnarray}
%I(\tau) &=& -\sigma A^{-a} B^{1-b} \frac{A_s}{A} \\
%J(\tau) &=& -\sigma A^{-a} B^{1-b} \frac{B_s}{B}
%\end{eqnarray}
%Eq \ref{Eq:AB} reduces to
%\begin{equation}
%w_{\tau} = D_y(w) -I(\tau) y w_y + J(\tau) w
%\end{equation}
%
This is the ``co-exploding" or ``co-collapsing" equation which, for
self-similar problems, is analogous to the ``co-traveling" equation for
translationally invariant ones.
Two additional constraints, frequently called {\it pinning
conditions}, are needed to find both $A(\tau)$ and $B(\tau)$.
There is some latitude in the selection of these conditions -- different
conditions correspond to different ways of ``moving along with the
solution across scales" before Eq. (\ref{Eq:AB}) reaches steady state,
although all appropriate pinnings give the
same self-similar shape and exponents \cite{Kurt}.
In the spirit of Rowley and Marsden, we proposed in \cite{Betelu} that such
conditions can be constructed by imposing relationships between
the solution and (essentially arbitrary)  template functions.
%

%{\bf ***}
Since there are two degrees of freedom (``amplitude'' and ``width'')
we must impose two pinning conditions.  Once these have been imposed,
we can use Eq. (\ref{Eq:AB}) to solve for $A(\tau)$ and $B(\tau)$: it is
actually possible to eliminate the $A_{\tau}/A$ and
$B_{\tau}/B$ terms in Eq. (\ref{Eq:AB}) to end up with a
``co-exploding" PDE that we called in \cite{Betelu} MN-dynamics.
When (and if) the solution of this PDE approaches an asymptotic steady
state, we can compare coefficients in Eqs. (\ref{Eq:ode}) and (\ref{Eq:AB})
to find:
\begin{equation}
\frac{\alpha}{\beta} = \frac{\lim_{\tau\rightarrow \infty} A_{\tau}/A}
     {\lim_{\tau\rightarrow \infty} B_{\tau}/B}
\label{Eq:exp2}
\end{equation}
Eqs. (\ref{Eq:exp1}) and (\ref{Eq:exp2}) allow us to obtain the scaling
exponents $\alpha$ and $\beta$.
A more detailed discussion can be found in \cite{Kurt}; the approach can
be used to locate both types of self-similar solutions \cite{Barenblatt},
and indeed in \cite{Betelu} it was used to locate both the Barenblatt and
the Graveleau solutions of the porous medium equation.

In view of our illustrative example using molecular dynamics below, we
briefly study the very simple case of the 1-dimensional diffusion equation,
\begin{equation}
u_t = u_{xx}.
\label{Eq:1d}
\end{equation}
The operator in this case is $D_x(u) = u_{xx}$. From Eq. (\ref{eq2}), $b =
1$ and $a = -2$.
Eq. (\ref{Eq:exp1}) then yields $\alpha = 1/2$.
For self-similar solutions of the second kind, $\alpha$ and $\beta$ cannot be
computed {\it a priori}, and will
be found as part of the solution process.
For the two pinning conditions, we choose here to let $w$
satisfy the following relations
to two given template functions, $T_1(y)$ and $T_2(y)$ respectively,
\begin{eqnarray}
\int_{-\infty}^{+\infty} w(y, \tau)T_1(y)dy &=& 0 \label{Eq:orth} \\
\int_{-\infty}^{+\infty} w(y, \tau)T_2(y)dy &=& 1.
\label{Eq:norm}
\end{eqnarray}
These can be thought of as controlling the ``width'' and the
amplitude of the solution.
By multiplying Eq. (\ref{Eq:AB}) with each $T(y)$,
integrating and using the
above two equations, we can eliminate $A$ and $B$.
A discussion of important technical conditions,
having to do with the existence of these
integrals over infinite domains, and the appropriate solution spaces,
can be found in \cite{Beyn,Wayne}.
(An alternative approach is to require that the difference between the
solution of Eq. (\ref{Eq:AB}) and a template is minimized, as
discussed in \cite{Betelu,ClancyYannis}.)

For our simple diffusion example, we set $T_1(y)$ to be
\begin{equation}
T_1(y) = \left\{ \begin{array}{cc}
               1 & {\rm for~} |y|\leq 1/2 \\
          -1 & {\rm for~} |y|> 1/2.
               \end{array}
       \right.
\label{t1def}
\end{equation}
From Eqs. (\ref{Eq:AB}), (\ref{Eq:orth}), and (\ref{t1def})
we get an equation for $A_{\tau}/A$.
Since the subspace of solutions symmetric around $x=0$ is invariant,
restricting our search to symmetric solutions we find that
\begin{equation}
\frac{A_\tau}{A}  = - \frac{2w_y(0.5,\tau)}{w(0.5,\tau)}.
% if not symmetric
% \frac{A_\tau}{A} = \frac{2w_y(-0.5)-2w_y(0.5)}{w(-0.5)+w(0.5)}
\label{eqt1inf}
\end{equation}
%(We will start with a symmetric initial condition below, so $w$
%reamins symmetric for all time.  Even if we started with an
%unsymmetric condition, it will become symmetric at steady state, so
%the use of Eq. (\ref{eqt1inf}) is a valid way of computing the limiting
%steady-state solution.)
%
Similarly, if we set $T_2(y) = \delta(y)$ we get
\begin{equation}
\frac{B_\tau}{B} = w_{yy}(0,\tau).
% if not symmetric
% \frac{B_\tau}{B} = w_{yy}(0,\tau)-w_{\tau}(0, \tau)
\label{eqt2inf}
\end{equation}
Substituting these in Eq. (\ref{Eq:AB}) we get
\begin{equation}
w_{\tau}+w_{yy}(0, \tau)w+\frac{2w_y(0.5, \tau)}{w(0.5, \tau)} y w_y  = w_{yy}.
% if not symmetric
% w_{\tau}+(w_{yy}(0, \tau)-w_{\tau}(0, \tau))w+  \frac{2w_y(0.5)-2w_y(-0.5)}{w(-0.5)+w(0.5)}y w_y  = w_{yy}
\label{Eq:diff1}
\end{equation}

For an initial condition consistent with the constraints and the
symmetry we take
\begin{equation}
w(y, 0) = \left\{ \begin{array}{cc}
                  1 & {\rm for~} |y|\leq 1 \\
              0 & {\rm for~} |y|> 1.
               \end{array}
       \right.
\end{equation}
so that both Eqs. (\ref{Eq:orth}) and (\ref{Eq:norm}) are satisfied.
Finally, we numerically integrate Eq. (\ref{Eq:diff1}) to steady state,
i.e. $\tau\rightarrow \infty$, and evaluate the exponent, $\beta$ from
\begin{equation}
\beta = -\frac{\alpha w(0.5, +\infty)w_{yy}(0, +\infty)}{2 w_y(0.5, +\infty)}.
\end{equation}

Another choice for $T_2(y)$ would be $T_2(y) = 1$.
Physically it corresponds to conservation of mass, and the same
procedure leads to
\begin{equation}
w_{\tau} + \frac{2w_y(0.5, \tau)}{w(0.5, \tau)}(w + yw_y) =
w_{yy}.
\label{Eq:diff2}
\end{equation}
Without any further integration, it is clear that
$\beta = -\alpha = -1/2$.

\begin{figure}[htbp]
\centering
\leavevmode
\includegraphics[height = 2.6in]{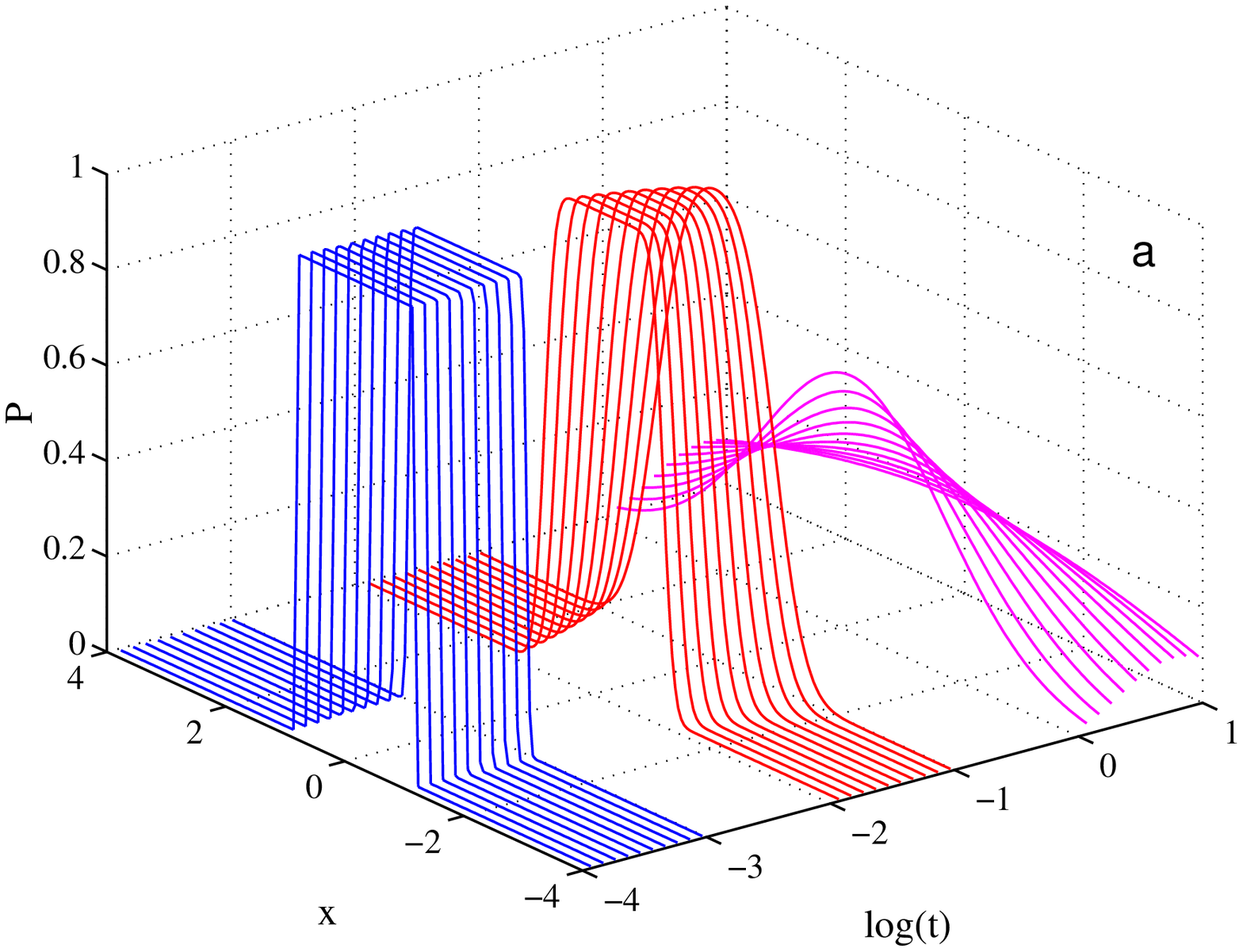}
\includegraphics[height = 2.6in]{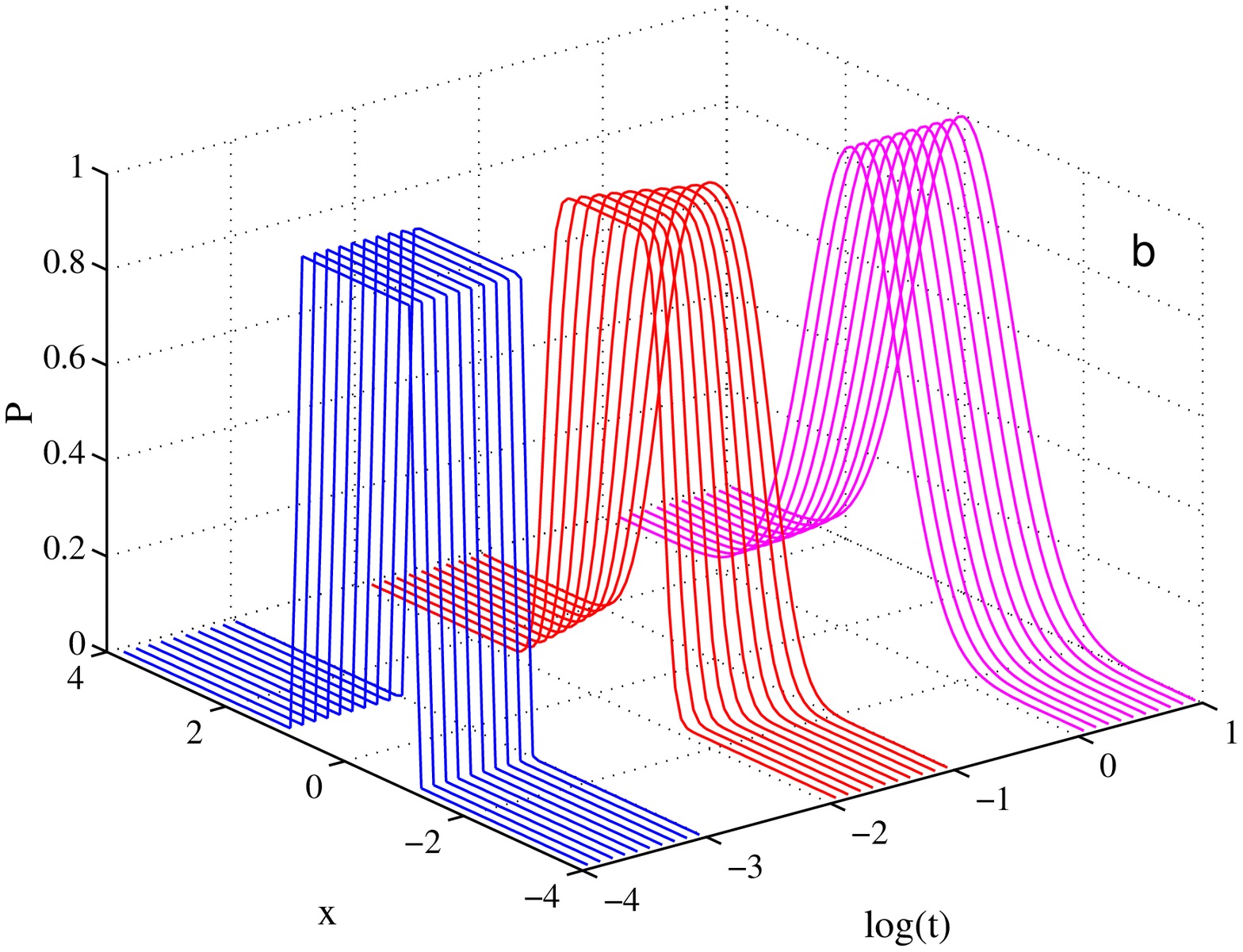}
\includegraphics[height = 2.6in]{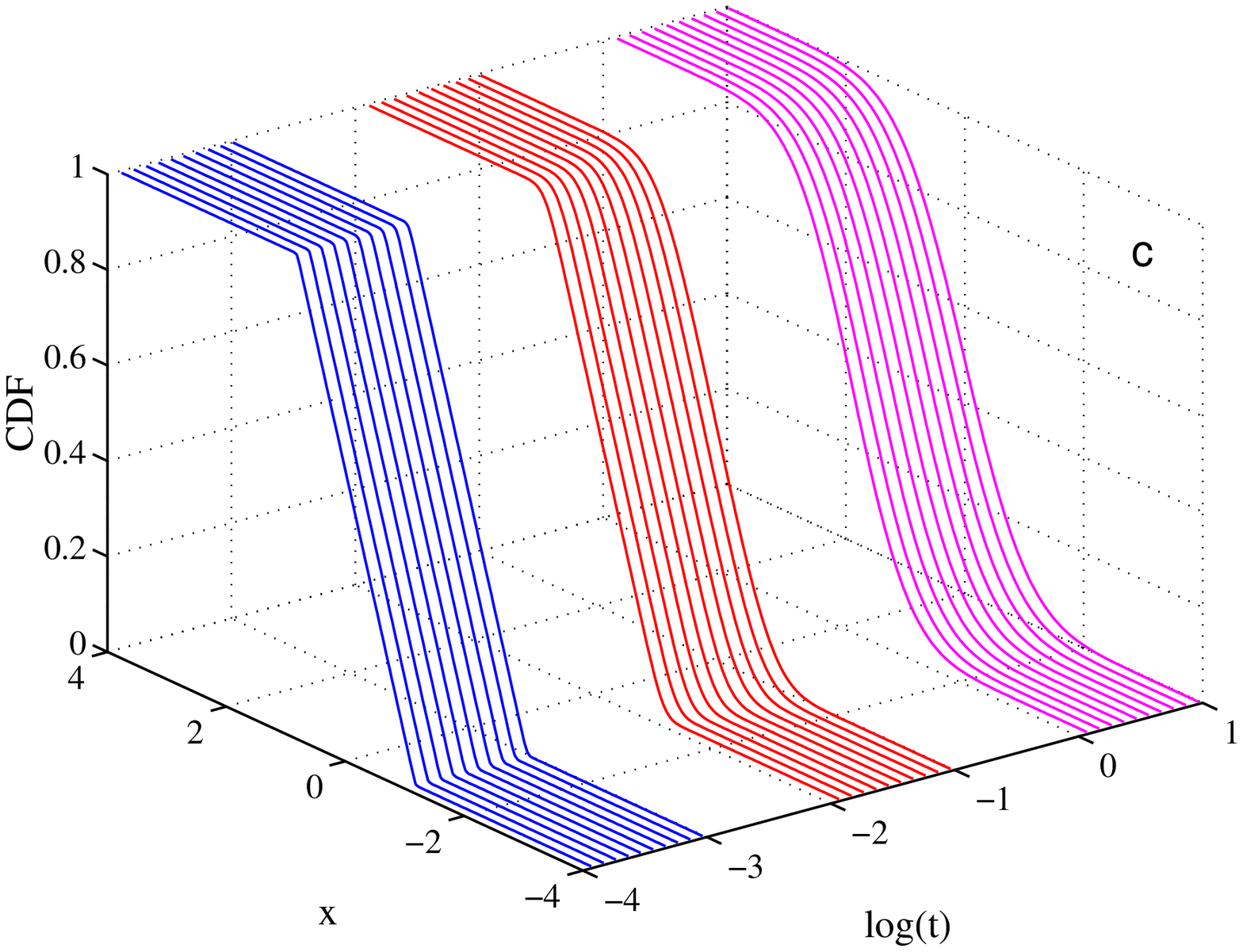}
\caption{ a) Time evolution of an initial rectangular density profile
by the 1d diffusion equation, Eq. (\ref{Eq:1d}). b) Dynamically
renormalized evolution of the same shape using Eq. (\ref{Eq:diff1}). c)
Cumulative distribution function representation of (b).}
\label{fig1}
\end{figure}

The above methodology evolves the differential equation in a
dynamically rescaled time and space frame.
Figure \ref{fig1}a shows the solution of the diffusion equation (in
the central portion of a large domain) for the initial condition
above; as time progresses we know that the solution decays, but it
also asymptotically approaches a (self-similarly decaying) Gaussian.
(For clarity, the solution is shown over several small blocks of time,
alhough the equation was also integrated over the ``spaces'' in
the figure.)
Figure \ref{fig1}b shows the same evolution {\it in a rescaled frame}
using Eq. (\ref{Eq:diff1}); the decay has now been removed through the
template-based rescaling, and one only sees the transient approach to
the self-similar shape (the Gaussian) consistent with the
template-based pinning conditions.
Figure \ref{fig1}c shows the same results in terms of {\it cumulative
material density} and not the density itself; as we will discuss
below, it is numerically more convenient in particle based simulation
codes to work with the {\it cumulative distribution function} rather
than the particle distribution function itself.
It is interesting to consider the case in which we have a so-called
``legacy code'' -- a code that evolves the original equation and which
we can run but cannot modify.
The so-called {\it numerical analysis of legacy codes} allows us to
transform a direct legacy dynamic simulator, by ```wrapping" a
computational superstructure around it, into a code capable of
performing a different set of tasks, for which the legacy simulator
was not designed.
In the dynamic renormalization case, we will compute self-similar
solutions by evolving in physical variables and rescaling the results,
as opposed to first obtaining and then evolving the rescaled equation.
Our approach is a discrete-time approach (see also
\cite{Olof,ChenGoldenfeld}); pioneering work on dynamic
renormalization, especially in the context of the Nonlinear
Schr\"{o}dinger Equation can be found in
\cite{MPSS,LePSS,LePSS2,LaPSS,KL,SS,LaPSSW,RW,ZS,KSZ}.

{\noindent \bf \large 3. Timestepping for coarse dynamic renormalization}

\begin{figure}[htbp]
\centering
\leavevmode
\includegraphics[height = 2in]{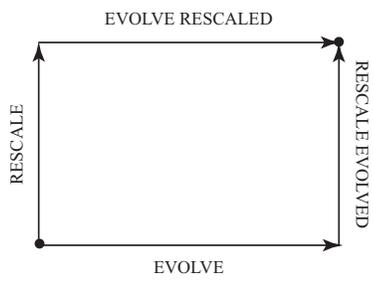}
\caption{Rescaling the finite time direct simulation
commutes with the dynamic renormalization flow.}
\label{fig2}
\end{figure}

As we discussed above, one can use a direct simulator of the original
equation to converge computationally to a (stable) self-similar
solution as follows.
Starting with an initial profile, one evolves forward for a finite
time; one then uses the template conditions to rescale the space
variable for the final profile.
The rescaled profile is then given as an initial condition to the
original equation, which is again evolved over finite time;
the space variable is rescaled, and the procedure is iterated
until the shape converges to a member of the family of self-similar
shapes.
The idea here is that rescaling, and then evolving the rescaled equation
for a finite time, commutes with evolving in physical space and then
rescaling the result (see Figure \ref{fig2}).
Also, while stable self-similar solutions can be found by such
dynamic rescaling and forward integration, they can also be found
  -- and so can unstable self-similar solutions --  through fixed point
algorithms (like Newton-Raphson).
Indeed, if we call $\Phi_{T}(w(y))$ the result of integrating
the rescaled equation with initial condition $w(y)$ over time $T$,
the self-similar solution satisfies
\begin{equation}
w(y) - \Phi_{T}(w(y)) = 0.
\end{equation}
We have shown in the past how matrix-free iterative linear algebra
techniques \cite{Kelley} can be used to converge to solutions of
such problems even when the only available tool is a subroutine that
numerically computes $\Phi_{T}(w(y))$.
The original inspiration for this work was the so-called Recursive
Projection Method (RPM) of Shroff and Keller \cite{ShroffKeller}, who used
this subroutine (the timestepper) and a computational superstructure
(the RPM wrapper) to construct a fixed point algorithm.
Under appropriate conditions this algorithm accelerates the
computation of stable fixed points and also stabilizes the computation
of dynamically unstable ones.
In effect, timestepping combined with matrix-free techniques ``fools"
a dynamic simulator into becoming a fixed point solver.
It is important to note in this entire exposition that we have
essentially ignored here the role of the boundary conditions for the
original and the rescaled equation, assuming we are solving in
sufficiently large domains and for sufficiently short times; it is
conceivable that weighted Sobolev spaces must be used in order to
avoid spurious numerical oscillations \cite{Wayne}.
This is an important issue which must be studied carefully; yet, as we
will see, for our simple problem this does not create major
computational difficulties.

\begin{figure}[htbp]
\centering
\leavevmode
\includegraphics[height = 5in]{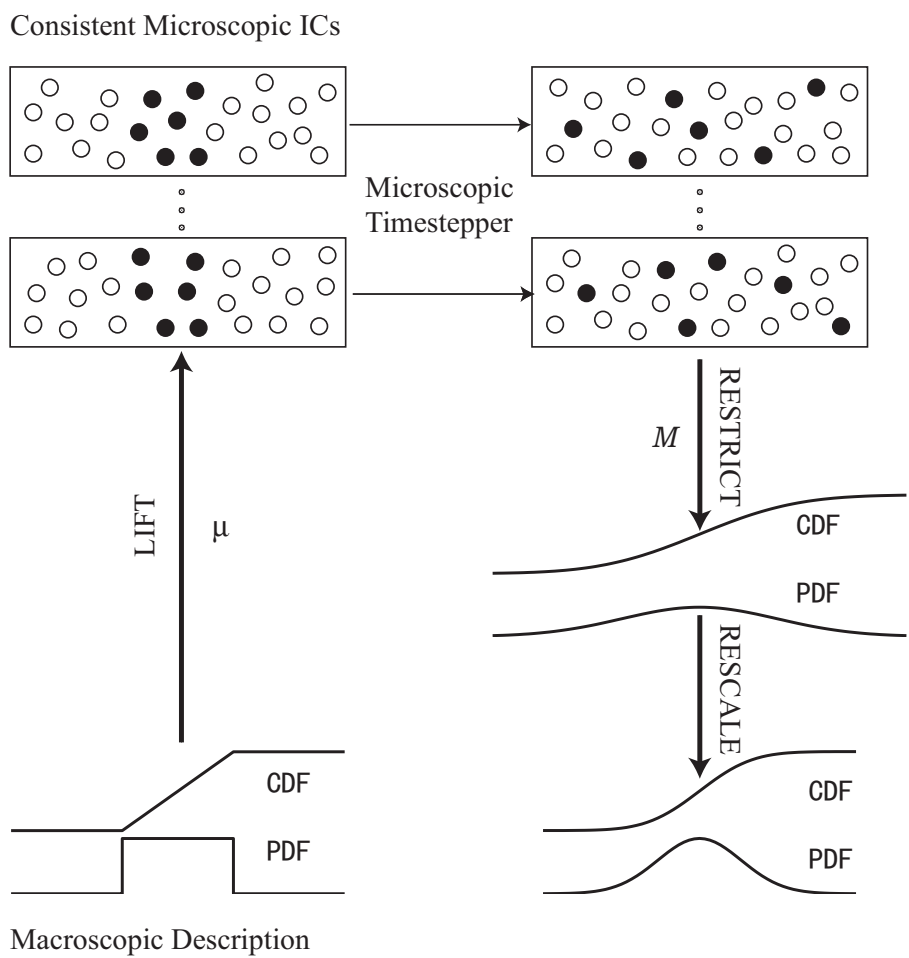}
\caption{Schematic view of the coarse dynamically renormalized timestepper.}
\label{fig3}
\end{figure}

We now return to the premise of our introduction: we have a
microscopic code (e.g. molecular dynamics, evolving a distribution of
molecules); yet we believe that the coarse-grained, macroscopic
behavior {\it of the statistics of the simulation} satisfies a macroscopic
equation that possesses self-similar solutions.
We will find these solutions through what we will call the ``coarse
timestepper", which we have extensively discussed in
\cite{TQK00,GKT02,MMK02,SGK03,LKG03,GLK03,GHK03,HK03,LB}, and which
-- for the case of dynamic renormalization -- is illustrated in Figure
\ref{fig3}.
This ``coarse dynamically renormalized timestepper" consists of the
following elements:

\begin{enumerate}
\item Choose the statistics of interest for describing the coarse-grained
behavior of the system and an appropriate representation for them.
In this case the concentration profile in one space dimension is the
appropriate macroscopic observable; it is the zeroth moment of the
distribution of molecules over velocities (and over the second, ``thin"
dimension).
It is more convenient to use the particle instantaneous Cumulative
Distribution Function; assuming it is smooth enough, one can use a
low-dimensional description of it based on the first few of an appropriate
sequence of orthogonal
polynomials \cite{Gear,Sima}.
We will call this the macroscopic description, ${\bf u}$. These
choices determine a restriction operator, $\restrict$, from the
microscopic-level description, ${\bf U}$ (particle coordinates) to the
macroscopic description: ${\bf u} = \restrict{\bf U}$.
\item Choose an appropriate {\it lifting} operator, $\lift$, from the
macroscopic description, ${\bf u}$, to the microscopic description,
${\bf U}$.  In our case we make random particle assignments consistent
with the macroscopic CDF.
$\lift$ should have the property that $\restrict\lift$ is the identity
($\restrict\lift=I$). In other words, lifting from the macroscopic to
the microscopic and then restricting (projecting) down again should
have no effect (except round-off).
\item From an initial value at the microscopic level, ${\bf U}(t_0)$,
run the microscopic simulator (the fine timestepper) for a (relatively
short) macroscopic reporting horizon $T$ generating the values ${\bf
U}(T)$. We may have to repeat this for several microscopic initial
conditions ${\bf U}_i(t_0)$, consistent with the same macroscopic one
${\bf u}(t_0)$, for variance reduction purposes.
\item Obtain the restriction ${\bf u}(T) = \restrict{\bf U}(T)$
(the average restriction, in the case of many copies).
\item Rescale ${\bf u}(T)$ to obtain ${\bf u_R(T)}$ (using the
template conditions).
\item Lift ${\bf u_R}(T)$ to get a new consistent microscopic ${\bf U}
= \lift{\bf u_R}(T)$ and use it as a new starting value for repeating
steps 3 to 6.
\end{enumerate}

Since the diffusion equation has a stable self-similar solution, one
can simply repeat the above procedure and observe the approach of the
{\it statistics} of the molecular description to the Gaussian; the
repeated dynamic coarse rescaling helps avoid the continuous decay of
the direct simulation towards zero.
Alternatively, as we will show, one can use coarse fixed point
algorithms (such as Newton-Raphson) to converge iteratively to the fixed
point of the coarse rescaled timestepper (as opposed to repeatedly
integrating and rescaling).
Finally, while this is a problem where the exponents are known {\it a
priori} through scaling, the formulas presented in Section 2 and the
rescaling history upon convergence to the self-similar solution helps
us estimate the limiting values of $A_{\tau}/A$ and
$B_{\tau}/B$.
This will then help recover, through Eqs. (\ref{Eq:exp1}) and
(\ref{Eq:exp2}) the self-similarity exponents for
type-2 self-similar solutions.

{\noindent \bf \large 4. The computational experiment}

In this Section we outline the data collection procedure from the
molecular dynamics simulation.
A standard Lennard-Jones potential was used, i.e.
\begin{equation}
V(r) = 4 \epsilon[(\sigma/r)^{12} - (\sigma/r)^6]
\end{equation}
with cutoff distance $2.5\sigma$.
In what follows all results will be given in reduced units, i.e.,
length in units of $\sigma$, time in units of $(m
\sigma^2/\epsilon)^{1/2}$.
The simulation was performed in a quasi-1d box with size $400 \times
10$, and periodic boundary conditions in both $x$ and $y$
directions.
We use $T = 1.0$, $\rho = 0.5$.
The domain for such a simulation time was
``large enough" to appear infinite over the time of our simulation.

First, the system is allowed to evolve to thermal equilibrium
(as evidenced by stationarity of the average kinetic energy of the particles).
%{\bf **** WHAT IS MEANT BY THERMAL EQUILIBRIUM HERE?  PROBABLY A CHEMICAL
%ENGINEER UNDERSTANDS WHAT YOU MEAN, BUT TO ME AND OTHERS IT MEANS
%EVOLVING UNTIL THERE IS NO FURTHER TEMPERATURE CHANGE.  I THINK SOME
%ADDITIONAL WORDS ARE NEEDED (LOCAL?)}
After some simulation time we record our first equilibrium configuration,
and ``reset the clock" to $t=0$.
Subsequent equilibrium configurations are recorded at later times.
To model the diffusion process, a fraction of the particles were
colored according to prescribed distributions (consistent with given
density profiles), and their positions in the later configurations
were tracked.
%

%{\bf **** REWRITE}

We are going to work with a single-variable cumulative distribution
function (CDF) \cite{Gear}.
Suppose that $N$ particles are colored, and the $i$-th particle is in
position $x_i$.
Given all the colored particle positions, we can immediately obtain
their empirical CDF by sorting them, that is, relabeling them so that
$x_i \ge x_{i-1}, ~0 < i \le N$,  and plotting $x_i$ vs
$p_i = (i-0.5)/N)$.
We use the CDF rather than the density function because it is trivial
to generate the empirical CDF while it is computationally difficult
and error-prone to compute the empirical density function which is the
(ill-defined) derivative of the CDF.

Using only the $x$ coordinate of particle positions is justified by
our quasi-1d simulation box.
In fact, it is easier to work with the inverse CDF, or ICDF, $Q(p)$.
It gives the the $x$ coordinate of a given particle position, $i$, and
it can be read off directly as $Q(p(i))$.
Such continuous ICDFs can then be microscopically approximated by
coloring the particle with the nearest $x$-coordinate value in the
simulation box.
Using the data (positions of colored particles) from later time
configurations, we can estimate the dynamics of the CDF (or ICDF) evolution.

In the first numerical experiments, we ran the systems for short
bursts of time, then applied templates to rescale the ICDF.
Because the number of colored particles we used is constant, mass is
automatically conserved, so that only one additional template, $T_1$, is
needed for dynamic rescaling.
%
%Use of the ICDF permits the use of a greater range of templates.
%
For this second template, we computed the slope of the center
20\% of the colored particles by linear least squares, and normalized
it to a fixed value.
This implied a simple rescaling of the $x$
coordinates of the computed colored particle positions.
These
rescaled $x$ values were then used to ``lift" the rescaled ICDF:
we selected the closest particles to these coordinates in
the full set, and colored them for a further short burst of computation.
Using this ``evolve--restrict--rescale--lift" procedure repeatedly, the
correct functional form of the self-similar solution will arise
asymptotically: the shape of the coarse-grained description (the ICDF)
converges to the inverse of the integral of a Gaussian (see Figure 4.)

In the second numerical experiment, we used a Newton iteration to
converge to the stationary, dynamically renormalized solution -- that
is, the self-similar solution.
To do
this, we need to restrict the CDF to a finite
(preferably low-dimensional) approximate representation.
Orthogonal polynomials usually provide computationally simple approximations, but
unfortunately they are not useful for the CDF which has a possible
support from $-\infty$ to $+\infty$ (that is, the CDF is $c(x)$ where
$x$ has a potentially infinite range; $c(x)$ must be monotone
non-decreasing and must lie between $0$ and $1$).
This is a second, and more important, reason for using the ICDF,
$x(p)$ where $p$ lies in $[0,1]$.  This function can easily be
approximated by a low-degree polynomial over its range.

We wish to find a polynomial approximation, $Q(p)$, that approximates
the computed positions.  That is, on the finite set of points,
$\{x_i\}$, corresponding to $\{p_i\} = \{(i - 0.5)/N\}$, we would like
$Q(p_i) \approx x_i$.  This provides us our restriction of the
microscopic data.  Then we can evaluate $Q$ at any point in [0,1] in
the lifting process.  (Typically, we will evaluate it at each $p_i$ to
get an approximate $x_i$.)
We are interested in minimizing the error of the approximation only at
the arguments $p_i, i = 1, \cdots, N$.  If we use least squares
approximation with weights $w_i$ we thus want to minimize
\begin{equation}
\sum_{i=1}^N [x_i - Q(p_i)]^2 w_i.
\end{equation}

In the experiments we used unit weights, $w_i = 1$.  The best way to
do this computationally is to create a basis for the $N$-dimensional
vector space, $\phi_s, s = 1, \cdots, N$ such that the $s$-th basis
vector, $\phi_s$ has its $i$-th element defined as the value of an
$(s-1)$-st degree polynomial evaluated at $p_i$, that is, $\phi_{si} =
q_s(p_i).$ To make this basis set orthonormal, we require that
\begin{equation}
\langle \phi_s,\phi_t \rangle \equiv \sum_{i=1}^N q_s(p_i) q_t(p_i) =
\delta_{st}.
\end{equation}
The $q$ polynomials thus defined are scalar multiples of the
orthogonal polynomials defined in the usual way from the L2 norm over
the unit interval.

Since the steady state solution of $w$ is symmetric, its ICDF will be
an odd function of $p-0.5$.  For this reason, we chose orthogonal
polynomials on the finite set $\{-0.5 + (i-0.5)/N\}$.
For the purposes of the Newton iteration we considered a fifth
degree polynomial representation of the ICDF written as
\begin{equation}
Q(p) = c_1\phi_1(p-0.5) + c_3\phi_3(p-0.5) + c_5\phi_5(p-0.5).
\label{p5}
\end{equation}
The constancy of the number of colored particles provides one
template condition for rescaling.
In this example, the second template condition was applied by
requiring that $c_1$ be constant.
After each burst of microscopic simulation, the restriction to form
Eq. (\ref{p5}) was performed, and then the polynomial was divided by
$c_1$ to get a scaled polynomial with $c_1 = 1$.
The use of a small number of features of the solution (``observables'') for
the Newton iteration with timesteppers is justified when
the long-term dynamics of an evolution equation are low-dimensional
(i.e. lie on an attracting low-dimensional manifold parameterized by a
small number of ``modes'' or ``observables'', like the $c_i$ here).

We can perform variance reduction by using multiple copies.  In this
case, because the box is large enough and the problem is
translationally invariant, different parts of the box can be used for
different realizations of evolving CDFs (and each can be colored
differently so that they can be distinguished, as long as they are
far enough apart to be uncorrelated).
We take an average of several such realizations (typically 10, or, for
stationary point computations, 100).

{\noindent \bf \large 5. Results and discussion}

\begin{figure}[htbp]
\centering
\leavevmode
\includegraphics[height = 4.5in]{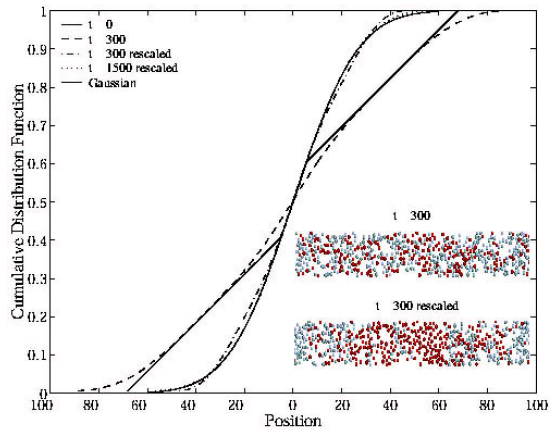}
\caption{Coarse evolution of the cumulative distribution function
using coarse renormalized timestepping, starting with a piecewise
linear CDF. The inset shows (top) a snapshot obtained around the
center of the computational domain after $t=300$, as well as (bottom)
the result of restricting, rescaling and lifting this snapshot.}
\label{fig4}
\end{figure}

In order to demonstrate how rescaling can accelerate convergence to
the self-similar shape we design a piecewise linear CDF with fewer
particles in the tail part, as shown in Fig. \ref{fig4}.
We evolve until $t=300$ and rescale, but the tail part of the rescaled
CDF is still significantly away from the Gaussian.
In the right lower corner of Figure \ref{fig4} snapshots of the
colored particles in the simulation box at $t=300$ and after rescaling
and lifting are shown.
It takes five repetitions of the procedure (evolving and rescaling)
until the rescaled CDF converges visibly to the Gaussian curve.

To reduce the effect of noise in estimating functional derivatives or
the fixed point algorithm, $200$ copies of the system at equilibrium
are let to run further until $t=1000$.
A perturbation as large as $3\%$ of the initial values of the coarse variables
(the $c_i$) is
necessary to ensure a meaningful finite difference estimate of the
$2\times 2$ Jacobian of the coarse self-similar fixed point
computation.
As shown in Table \ref{table1}, the fixed point values of $c_3$, $c_5$
are $c_3/c_1 \approx 0.18$, $c_5/c_1 \approx 0.075$, respectively.
%
%The corresponding values of a real Gaussian curve are
%$c_3/c_1 = 0.216214 & c_5/c_1 = 0.0790498$, respectively.
%

Comparing the fixed point solution shape (reconstructed based only
on the computed polynomial terms) with the Gaussian, the only
visual difference is at the tail part, where very few particles exist.
The actual solution we have found (the result of initializing
symmetrically with given values for the first three odd $c_i$ and zero
for the remaining $c_i$, evolving for the given time horizon, and
rescaling so that the first three odd $c_i$ have the same values) has
now acquired components in the remaining odd $c_i$, and is closer to
the Gaussian than its truncation.
Using molecular dynamics {\it constrained} on these three macroscopic
observables will give us a better estimate of the macroscopic fixed
point we located (see \cite{Gear2003} for further discussion).
Different basis sets for the approximation can be used if significant
information in the tails is not well captured this way.

The process we have described allows us to compute the shape of the
self similar solution.
If we also need the exponents $\alpha$ and $\beta$ in
Eq. (\ref{alpbet}) we have to know $a$ and $b$ defined in
Eq. (\ref{eq2}) and the value of $(\lim A_\tau/A)/(\lim B_\tau/B)$ so
that we can apply Eqs. (\ref{Eq:exp1}) and (\ref{Eq:exp2}).
When an equation is known, $a$ and $b$ are obtained by inspection.
When it is not known in closed form, ideas similar to those used in
\cite{LKG03} can be used to test the existence of self-similarity and
estimate $a$ and $b$; in particular, this will involve the microscopic
simulation for short periods to estimate $du/dt$ in Eq. (\ref{eq1})
from multiple initial conditions that are scaled (in amplitude and
space) versions of each other.

Once a coarse self-similar solution has been converged to, whether through
integration or Newton-Raphson of the appropriately renormalized coarse dynamics,
we can use again short simulations to estimate $A_t/A$ and $B_t/B$.
Starting a microscopic simulation at time $t$ and evolving for time $d$, the
relative values of $A$ and $B$,
that is, $A(t+d)/A(t)$ and $B(t+d)/B(t)$, can be obtained from the
scaling needed to re-impose the template conditions.
These lead to approximations of $A_t/A$ and $B_t/B$, from
which $\alpha/\beta$ can be estimated using Eq. (\ref{Eq:exp2}).

{\noindent \bf \large 6. Conclusions}

We have described a systematic approach for the computation of coarse
self-similar solutions in situations where the only available model is
a microscopic (in this case molecular dynamics) simulator.
The procedure uses short, appropriately initialized bursts of MD
simulation to estimate the coarse timestepper of the template-based
renormalization flow for the process.
%
%{\it by the way - please find for me the eigenvalues of this !!}.
% -0.95 and -1
In the case we studied here, a $2 \times 2$ numerical Jacobian for the
Newton iteration was sufficient; more generally,
matrix-free techniques can be combined with this coarse renormalized
timestepper to compute stable as well as unstable self-similar
solutions, compute ``finite times to blow-up" in the appropriate cases
as well as estimate the self-similarity exponents.
The stability of the self-similar solutions in the co-exploding
frame can be probed through the same coarse renormalized timestepper and
matrix-free eigenanalysis techniques.
It is also possible to combine the equation-free
algorithms presented here with the so-called
"gaptooth scheme" and "patch dynamics" \cite{GLK03,Samaey,Samaey2,GHK03}.
The microscopic computations are not performed across all
of space, but only in relatively small physical domains ("teeth"
separated by gaps and connected
through appropriate boundary conditions).
This approach exploits smoothness of the macroscopic
observables (e.g. particle density) in space as well as time, in order to
further reduce the microscopic simulation necessary to compute coarse
self-similar solutions.

It is important to notice that the results are valid over some scaling
regime, while for model equations they in principle hold over all
scales.
In the context of non-Newtonian fluid mechanics, such techniques could
assist the quantitative detection of self-similarity in phenomena such
as spreading \cite{Ansini,Arons2003}.
More ambitiously, one may envision the use of equation-free coarse
renormalization to study the self-similar behavior of the evolution of
spectra in turbulence studies \cite{Melander}.
Extensive research explores the onset of dynamic self-similarity as
operating parameters vary \cite{Siettos} as well as the study of
asymptotically self-similar solutions.
Self-similar solutions may explode or collapse in infinite time (as in
the case of the diffusion equation we studied here) or can have
finite-time blow-ups, in which the solution increasingly accelerates in
time.
Conversely it is possible to have self-similar solutions which {\it
progressively decelerate} in time; this may be relevant in the
description of glassy dynamics \cite{Glass_paper}, and it is
conceivable that extensions of the approach presented here may assist
in accelerating the computation of self-similar solutions that
progressively slow down.

\section*{Acknowledgments}
This work was partially supported by AFOSR (Dynamics and Control) as
well as an NSF/ITR grant. Discussions with Profs. P. Kevrekidis, C. Rowley,
D. G. Aronson, S. Betelu and Dr. K. Lust are gratefully acknowledged.

\newpage
%\bibliography{diffuse}

\newpage
{\bf Figure Captions}

{\it Fig. 1}
a) Time evolution of an initial rectangular density profile
by the 1d diffusion equation, Eq. (\ref{Eq:1d}). b) Dynamically
renormalized evolution of the same shape using Eq. (\ref{Eq:diff1}). c)
Cumulative distribution function representation of (b).

{\it Fig. 2}
Rescaling the finite time direct simulation
commutes with the dynamic renormalization flow.

{\it Fig. 3}
Schematic view of the coarse dynamically renormalized timestepper.

{\it Fig. 4}
Coarse evolution of the cumulative distribution function
using coarse renormalized timestepping, starting with a piecewise
linear CDF. The inset shows (top) a snapshot obtained around the
center of the computational domain after $t=300$, as well as (bottom)
the result of restricting, rescaling and lifting this snapshot.
\newpage

\begin{table}
\begin{center}
\caption{Dynamically renormalized fixed point computation using
Newton-Raphson.}
\begin{tabular}{cccc} \hline
iteration & $c_3/c_1$ & $c_5/c_1$ & $|\Delta c_3/c_1| + |\Delta c_5/c_1|$ \\ \hline
0 & 0 & 1 & 1.09626 \\ \hline
1 & 0.278266 & 0.156855  & 0.138427 \\ \hline
2 & 0.188089 & 0.0816683 & 0.01203 \\ \hline
3 & 0.176717 & 0.0745533 & 0.00207797 \\ \hline
4 & 0.178675 & 0.0747852 & 0.000377391 \\ \hline
\end{tabular}
\label{table1}
\end{center}
\end{table}

\newpage

\end{document}